\documentclass[aps,prb,preprint,superscriptaddress,amsmath,showpacs]{revtex4-1}
\usepackage{graphicx}
\begin{document}
\title{Condensation of collective charge ordering in Chromium}
\author{A. Singer}
\altaffiliation{ansinger@ucsd.edu}
\affiliation{Department of Physics, University of California, San Diego, La Jolla, California 92093, USA}
\author{M. Marsh}
\affiliation{Department of Physics, University of California, San Diego, La Jolla, California 92093, USA}
\author{S. Dietze}
\affiliation{Department of Physics, University of California, San Diego, La Jolla, California 92093, USA}
\author{V. Uhl\' \i\v r}
\affiliation{Center for Magnetic Recording Research, University of California, San Diego, La Jolla, California 92093, USA}
\author{Y. Li}
\affiliation{Advanced Photon Source, Argonne National Laboratory, Argonne, Illinois 60439, USA}
\author{D. A. Walko}
\affiliation{Advanced Photon Source, Argonne National Laboratory, Argonne, Illinois 60439, USA}
\author{E. M. Dufresne}
\affiliation{Advanced Photon Source, Argonne National Laboratory, Argonne, Illinois 60439, USA}
\author{G. Srajer}
\affiliation{Advanced Photon Source, Argonne National Laboratory, Argonne, Illinois 60439, USA}
\author{M. P. Cosgriff}
\affiliation{Department of Materials Science and Engineering, University of Wisconsin Madison, Madison,Wisconsin 53706, USA}
\author{P. G. Evans}
\affiliation{Department of Materials Science and Engineering, University of Wisconsin Madison, Madison,Wisconsin 53706, USA}
\author{E. E. Fullerton}
\affiliation{Center for Magnetic Recording Research, University of California, San Diego, La Jolla, California 92093, USA}
\author{O. G. Shpyrko}
\affiliation{Department of Physics, University of California, San Diego, La Jolla, California 92093, USA}
\begin{abstract}
Here we report on the dynamics of the structural order parameter in a chromium film using synchrotron radiation in response to photo-induced ultra-fast excitations. 
Following transient optical excitations the effective lattice temperature of the film rises close to the N\'{e}el temperature and the charge density wave (CDW) amplitude is reduced. 
Although we expect the electronic charge ordering to vanish shortly after the excitation we observe that the CDW is never completely disrupted, which is revealed by its unmodified period at elevated temperatures. 
We attribute the persistence of the CDW to the long-lived periodic lattice displacement in chromium. 
The long-term evolution shows that the CDW revives to its initial strength within 1 ns, which appears to behave in accordance with the temperature dependence in equilibrium. 
This study highlights the fundamental role of the lattice distortion in charge ordered systems and its impact on the re-condensation dynamics of the charge ordered state in strongly correlated materials.
\end{abstract}

\pacs{61.05.C-,63.20.kd,71.45.Lr,75.78.-n}
\maketitle

The subtle long-range order parameters of solids, including those induced by electronic and magnetic interactions, can have unusual dynamics due to the coupling of phenomena with vastly different timescales. 
Understanding and controlling the interaction among various degrees of freedom (spin, charge, and lattice) in strongly correlated systems across these timescales provides an important series of challenges and opportunities \cite{1}. 
The spectrum of timescales arises because electron-electron interaction is exceptionally fast, on the order of femtoseconds; optical pump-probe spectroscopy, for example, has demonstrated a breakdown of the charge-ordered insulating phase of manganites in less than 1 ps \cite{2}. 
Recent time- and angular-resolved photoelectron spectroscopy measurements confirmed these ultrafast time scales and have shown that the CDW in TiS$_2$ vanishes even faster, within the first 100 fs after the impact of an infrared (IR) laser pulse \cite{3}.
The lattice, however, is limited to far longer timescales characteristic of the propagation and damping of vibrational modes. In the ground state the interaction between electronic and lattice degrees of freedom plays a crucial role in a variety of phenomena such as antiferromagnetism \cite{4}, superconductivity \cite{5,6}, and colossal magnetoresistance \cite{7}, to name a few. 
Understanding the coupling of these degrees of freedom away from equilibrium is not only fundamental for basic science but also may drive new applications.

There is considerable interest in understanding the time scales involved in destroying and creating CDWs in various materials by accessing the temporal response of the photo-responsive degrees of freedom \cite{2,3,8,9}. 
Despite its importance, the dynamics of ionic ordering, the elastic component of the CDW, has gained less attention in the past. 
Recent experimental observations show melting of the ionic ordering in picoseconds and varying re-condensation timescales of microseconds in manganites \cite{2}, picoseconds in TiSe$_2$ \cite{8,10}, and even hundreds of picosecond in 4Hb-TaSe$_2$ \cite{9}. 
The way in which the dimensionality of the problem affects the timescale is debated \cite{9}, and the role of inherent inhomogeneities in these materials remains elusive \cite{11}. 
Moreover, recent theoretical \cite{12,13} and experimental studies \cite{9} indicate that Fermi-surface nesting, which is often used to describe the origin of CDW \cite{4,14,15}, may not be sufficient to describe the charge ordering and that electron-phonon coupling plays an indispensable role. 

Here we report on sub-nanosecond dynamics of ion charge ordering in chromium (Cr), a strongly correlated metal with a simple body-centered cubic structure that forms an antiferromagnetic spin-density wave (SDW) and CDW.
Chromium is a particularly good model as it lacks all of the disorder and inherent complexity present in other charge ordered systems \cite{7,11}.
We use ultra-fast IR laser pulses to excite the electronic subsystem in a Cr thin film and use x-ray diffraction to study the average lattice expansion (via the Bragg peak) and the amplitude and period of the CDW (via satellite peaks). 
Despite substantial excitation of electrons by the IR pulse (to electron temperatures well above the ordering temperature), the ionic charge ordering is present at all times following the excitation.
A reduction in charge ordering is observed, and can be linked to a rise in temperature, but rapidly recovers due to cooling by thermal conduction into the substrate.
 
%%%%%%%%%%%%%%%%%%%%%%%%%%%%%%%%%%%%%%%%%%%%%%%%%%%%%%%%%%%%%%%%%%%%%%%%%%%%%%%%%%%%%%%%%%
\begin{figure}[t]
\includegraphics[width=.7\columnwidth]{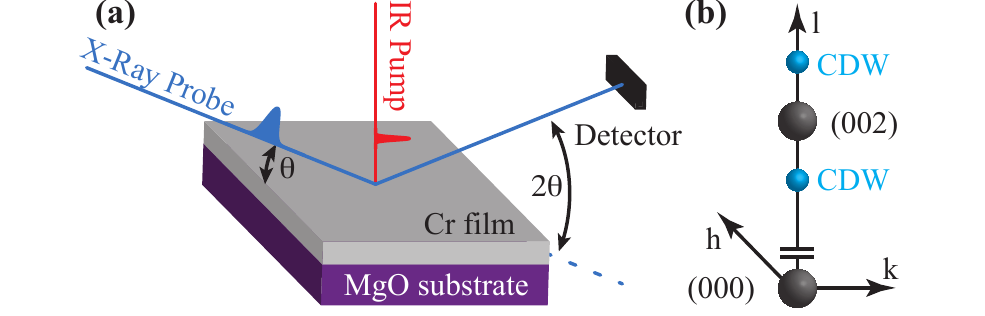}
\caption{(a) X-rays are incident on a thin chromium film and the diffracted intensity is recorded at the detector. The system is pumped using 60 fs IR pulses with an adjustable delay between IR and 100 ps x-ray pulses. (b) A schematic of reciprocal space of the chromium lattice around the (002) Bragg peak with the presence of a CDW. 
}
\label{fig:1}
\end{figure}
%%%%%%%%%%%%%%%%%%%%%%%%%%%%%%%%%%%%%%%%%%%%%%%%%%%%%%%%%%%%%%%%%%%%%%%%%%%%%%%%%%%%%%%%%%

The system under study was a Cr thin film deposited onto single-crystal MgO(001) substrate using magnetron sputtering \cite{16} at a substrate temperature of 800$^{\circ}$ C.
Short pulses (60 fs pulse duration) of IR laser radiation were used to initiate dynamics in the film. 
The IR laser was incident normal to the film surface (see Fig. \ref{fig:1}(a)) and had a wavelength of 800 nm, a repetition rate of 5 kHz, a spot size of about 0.5 mm diameter full width at half maximum (FWHM), and an energy of 80 μJ per pulse.
The pulse energy was adjusted to achieve significant excitation of the system yet avoid damage.
The film thickness was selected to match the optical skin depth of chromium, which is about 30 nm at a wavelength of 800 nm \cite{17}.
The temporal evolution of the structure in the film was studied with synchrotron radiation at the Advanced Photon Source, Argonne National Laboratory, utilizing x-ray diffraction scans in stroboscopic mode.
X-ray pulses had a photon energy of 10.5 keV, a pulse duration of 100 ps, and were focused to a spot size of 100 $\mu$m FWHM yielding a footprint well below the IR laser spot.
The time delay between IR laser and x-ray pulses was varied between -100 ps and 1 ns.
An avalanche photodiode (APD) was used to record the signal from individual synchrotron pulses.

Below the N\'{e}el temperature (311 K in bulk) \cite{4} Cr is antiferromagnetic and exhibits an itinerant SDW \cite{4,15}, accompanied by a CDW.
The CDW can be considered as a harmonic of the SDW \cite{4}. 
In the thin-film geometry the CDW has a wave vector pointing normal to the film surface along the (001) direction and is incommensurate with the lattice. 
Therefore the reciprocal space accessible in x-ray diffraction not only contains bright Bragg peaks due to the crystallinity of the film but also shows satellites due to the periodic lattice distortion associated with the charge ordering to both sides of the Bragg peak, separated by the inverse wavelength of the CDW (see Fig. \ref{fig:1}(b)). 
In bulk Cr the wavevectors of the satellites depend on the temperature and the CDW wavelength varies between 3 and 4 nm.
In Cr thin films, the CDW is pinned by the interfaces yielding discrete wave periods that are hysteretic with temperature \cite{18,19,20}.

%%%%%%%%%%%%%%%%%%%%%%%%%%%%%%%%%%%%%%%%%%%%%%%%%%%%%%%%%%%%%%%%%%%%%%%%%%%%%%%%%%%%%%%%%%
\begin{figure}[t]
\includegraphics[width=.7\columnwidth]{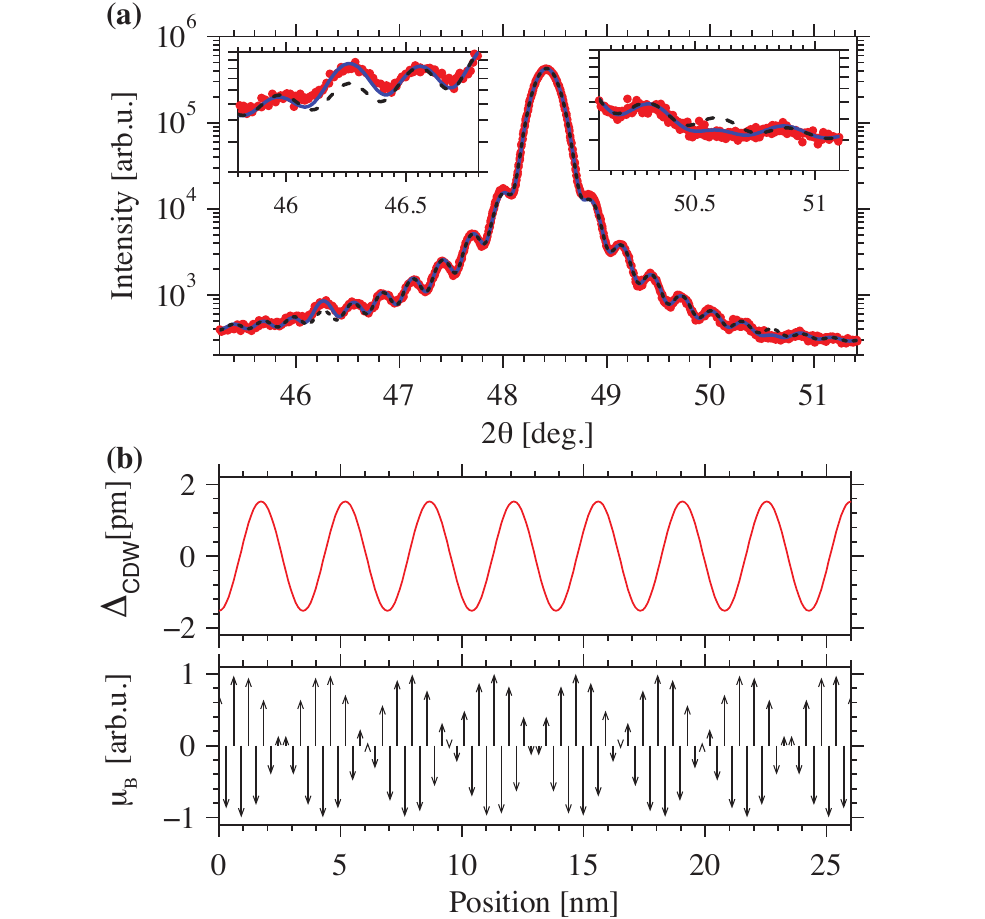}
\caption{(a) X-ray diffraction scan of the Cr film cooled to 20 K (red circles) and theoretical fits with CDW (blue solid line) and without CDW (black dashed line) taken into account. Insets show enlarged regions of the diffraction scan. (b) Displacement $\Delta_\textrm{CDW}$ (top) consistent with the x-ray diffraction data in (a). The magnetic moment $\mu_\textrm{B}$ (bottom) was inferred from the phase relation between the CDW and SDW \cite{22}. The interface with the substrate is at position 0. 
}
\label{fig:2}
\end{figure}
%%%%%%%%%%%%%%%%%%%%%%%%%%%%%%%%%%%%%%%%%%%%%%%%%%%%%%%%%%%%%%%%%%%%%%%%%%%%%%%%%%%%%%%%%%
An x-ray diffraction scan of the crystalline chromium film at 20 K is presented in Fig. \ref{fig:2}(a). 
It shows the (002) Bragg peak at 2$\theta$=48.4$^{\circ}$ and interference fringes due to the finite thickness of the film. 
The excellent visibility of these fringes over the whole scan range indicates high crystal quality, homogeneous thickness of the film across the x-ray beam area, and a small surface roughness ($\sim$1 nm). 
Fitting of the diffraction profile yields a film thickness of 25.6$\pm$0.3 nm (see Appendix \ref{A},\ref{B} for details).
The purely structural model fails at 2$\theta$=46.3$^{\circ}$ and 50.6$^{\circ}$, where satellite peaks due to the CDW are expected (see inset in Fig. \ref{fig:2}(a)). 
Surprisingly, the CDW does not lead to an increase in the diffracted intensity at both positions, as expected from previous studies \cite{4,15}. 
We find that the increase (decrease) of intensity at 2$\theta$=46.3$^{\circ}$ (2$\theta$=50.6$^{\circ}$) is due to constructive (destructive) interference between the thickness fringes and the CDW peaks (see Fig. \ref{fig:2}(a) and Appendix \ref{A}). 
The coincidence between the positions of the satellites and thickness fringes further indicates pinning of the CDW at the film boundaries, as expected, and a CDW period of 3.4 nm at 20 K. 

The interference effect allows us to determine not only the amplitude of the ion displacement associated with the CDW but also its phase with respect to the film interfaces. 
For our sample we find that the lattice displacement has 7.5 periods and antinodes at the film boundaries with negative (positive) displacement at the substrate (air) interface (see Fig. \ref{fig:2}(b)). 
The amplitude of the displacement has been determined to be 0.5$\pm$0.2\% of the lattice parameter, slightly higher than the bulk value observed for higher temperatures \cite{21}. 
Using the known phase relations between the CDW and SDW \cite{22} we can determine the position of the SDW (Fig. \ref{fig:2}(b)). 
Our results suggest the SDW antinodes are pinned at both interfaces  
\footnote{Theoretical (in Fe/Cr/Air) \cite{15} and experimental (in CrMn/Cr/CrMn) \cite{18} studies showed that the half-period of the SDW close to an interface can be larger than the period further away from the thin film boundaries, which is consistent with our observation.}.

\begin{figure}[t]
\includegraphics[width=.7\columnwidth]{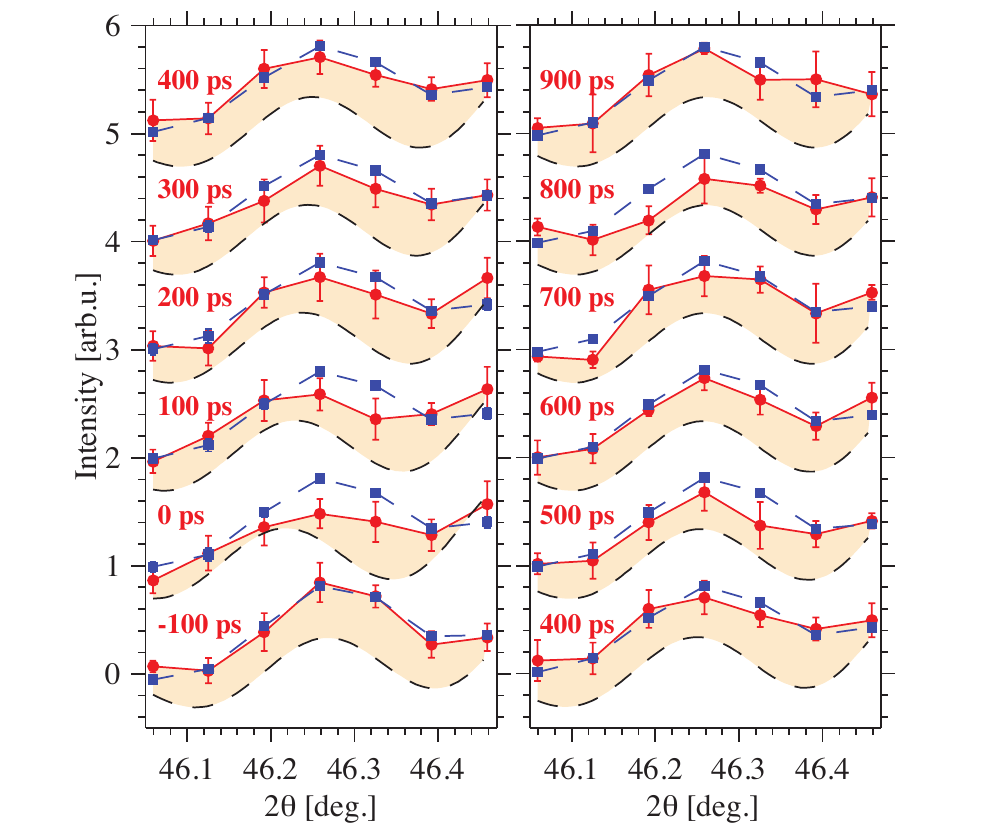}
\caption{X-ray diffraction scans for different time delays of the exited (laser on, red circles) and ground state (laser off, blue squares). Data for different delays are shifted vertically for better visibility. The black dashed line shows the intensity without CDW obtained from the fit in Fig. \ref{fig:2}(a) and corrected for the effective temperature (see below). The shaded region indicates the area corresponding to the CDW for different delay times.
}
\label{fig:3}
\end{figure}

To map the time evolution of the CDW after the impact of the excitation pulses we collected x-ray diffraction scans in the range around the CDW satellite at 2$\theta$=46.3$^{\circ}$ as a function of the time delay (see Fig. \ref{fig:3}).
Five to seven scans for each time delay were recorded (total measurement time about 60 hours) and the uncertainties in Fig. \ref{fig:3} show the deviations between these measurements.
X-ray data with ''laser on'' and ''laser off'' were collected simultaneously
\footnote{The APD was electronically gated to collect the “laser on” signal (x-rays arriving immediately after laser excitation). 
Due to the low repetition rate of the IR laser compared with the synchrotron source only 0.1\% of x-ray pulses are used to measure the excited state, while all other pulses scatter off the unpumped system.
The signal from all x-ray pulses was considered as the “laser off” signal.}.  
It is clearly seen that the intensity of the satellite peak scattered from the unpumped system (“laser off”) is quite consistent in all measurements, which shows that there were no instabilities or radiation damage during the measurement. 
The data taken with “laser on” for negative time delays is also consistent with the signal from the unpumped system (“laser off”).
%%%%%%%%%%%%%%%%%%%%%%%%%%%%%%%%%%%%%%%%%%%%%%%%%%%%%%%%%%%%%%%%%%%%%%%%%%%%%%%%%%%%%%%%%%
%%%%%%%%%%%%%%%%%%%%%%%%%%%%%%%%%%%%%%%%%%%%%%%%%%%%%%%%%%%%%%%%%%%%%%%%%%%%%%%%%%%%%%%%%%

The measurements for larger time delays reveal the dynamics of the CDW as a response to the IR laser pulse excitation. 
At 0 ps, where the IR pulse arrives just prior to the x-ray pulse, the diffracted intensity is weaker, which indicates a lower degree of charge ordering in the film. 
A comparison with the intensity expected without the presence of the CDW at 20 K (see Fig. \ref{fig:2}(a)) shows, however, that the elastic component of the CDW does not vanish and has a significant presence even shortly after the IR excitation pulse (Fig. \ref{fig:3}). 
The diffracted intensity increases for larger time delays and the CDW recovers close to its initial state after 900 ps. Note that for 0 ps the peak is smeared due to the inability of the relatively long x-ray pulse to capture fast dynamics (lattice expansion) that occur immediately after the IR pulse. 

\begin{figure}[t]
\includegraphics[width=.7\columnwidth]{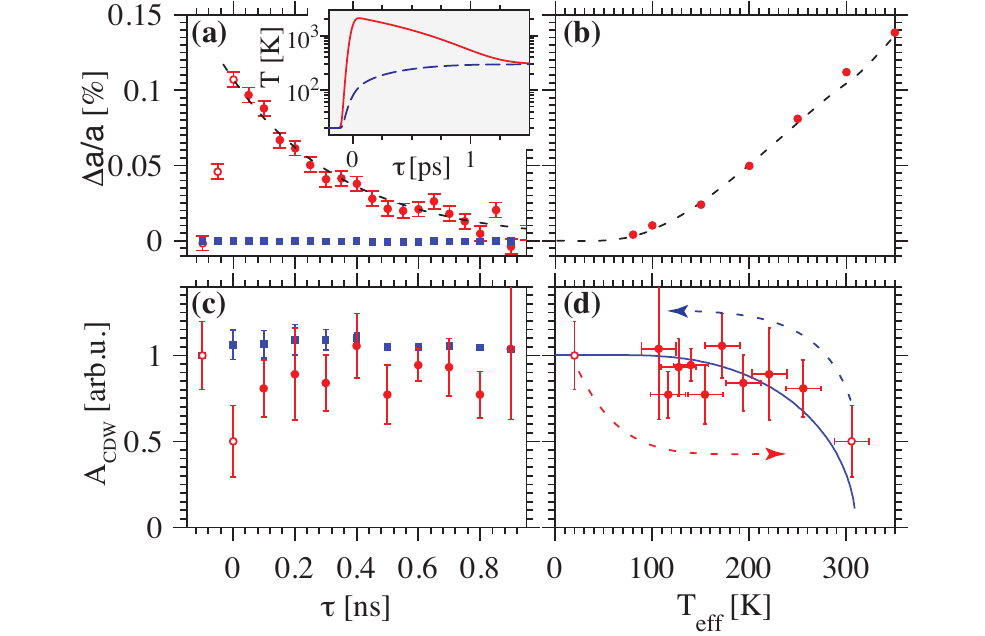}
\caption{(a) Relative shift $\Delta a/a$ of the average lattice constant in response to the IR laser pulse (red circles, filled circles represent delay times of 100 ps and above) and theoretical fit (black dashed line). 
(Inset) Calculation of the electron (red solid line) and lattice (blue dashed line) temperature (see text for details). 
(b) Thermal expansion coefficient of Cr along (001) in our sample (red circles) and bulk \cite{24} (black dashed line). 
(c) Amplitude $A_\textrm{CDW}$ of the CDW measured as a function of the time delay with laser on (red circles) and laser off (blue squares). 
(d) $A_\textrm{CDW}$ as a function of the effective temperature measured (red circles) and calculated within the BCS theory for a N\'{e}el temperature of 311 K (blue line). Note the higher sampling of $\tau$ in (a) as compared with (c).}
\label{fig:4}
\end{figure}

We used the relative Bragg peak shift as a thermometer for the effective temperature of the lattice, $T_\textrm{eff}$. 
At each time delay this shift was determined by measuring the x-ray diffraction signal at a fixed 2$\theta$ angle of 48.27$^{\circ}$, just on the slope of the Bragg peak. 
In the present experiment the lattice expanded by about 0.1\% shortly after the IR pulse at 0 ps (Fig \ref{fig:4}(a)). 
By measuring diffraction scans around 2$\theta$=48$^{\circ}$ we excluded significant broadening of the Bragg peak in response to heating. 
A theoretical fit to the data revealed a relaxation time of 370$\pm$40 ps (see Appendix \ref{C} for details). 
The signal recorded simultaneously from the unperturbed system is also shown in Fig. \ref{fig:4}(a) and is constant, as expected. 
We measured the thermal expansion coefficient on the same film using our lab source in equilibrium conditions (see Fig. \ref{fig:4}(b)). 
A comparison between Figs. \ref{fig:4}(a) and \ref{fig:4}(b) shows that the film was heated up to about 300 K at 0 ps 
\footnote{We have assumed that in the pump-probe experiment the film can expand in all directions equally due to structural inhomogeneities in the film and both measurements are comparable (see Appendix \ref{D} for a scanning electron microscope image of the film surface).}, which is close to the N\'{e}el temperature of bulk Cr \cite{4}. 

The diffraction data from Fig. \ref{fig:3} was used to calculate the height of the CDW satellite peak for each time delay (see Fig. \ref{fig:4}(c)) and effective temperature (see Fig. \ref{fig:4}(d)). 
This height was determined as the area of the shaded region in Fig. \ref{fig:3}, which corresponds to the difference between the measured data and the simulated Bragg peak intensity from a displacement-free lattice (Fig. \ref{fig:2}(a)). 
The latter was corrected for thermal expansion due to the effective temperature. 
Figure \ref{fig:4}(c) shows that the amplitude of the CDW is reduced to about 50\% of its initial value at -100 ps and increases at later times. The uncertainties in Figs. \ref{fig:4}(c) and \ref{fig:4}(d) were determined as averages over the uncertainties in the scattered intensities shown in Fig. \ref{fig:3}. 
The CDW amplitude can be well-described using BCS theory \cite{5,23} and Fig. \ref{fig:4}(d) shows the behavior predicted by the theory in bulk chromium being in equilibrium at a every temperature \footnote{Typically the CDW satellite peak intensity therefore scales with the square of the ordered magnetic moment of the spins in the SDW \cite{4}. 
We show (see Appendix \ref{A} for details) that due to interference, here, the CDW contribution to the measured satellite intensity scales linearly with the CDW amplitude and thus with the SDW intensity.} . 
For all measured points the data appears to agree with this theoretical interpretation, indicating that for time scales probed in our experiment the electrons and lattice are coupled and in equilibrium.

We can draw valuable conclusions from the existence of the CDW at 0 ps, although we have not measured the CDW dynamics on time scales shorter than the x-ray pulses (100 ps). 
In particular, we utilize the fact that the CDW period in Cr is strongly temperature dependent and shows a hysteretic behavior in thin films \cite{4,18} and expect it to be larger, if the CDW was completely disrupted by the IR excitation and re-emerged at elevated temperatures. 
A CDW with a different periodicity would manifest itself in scattering at a different angle, on top of a different thickness fringe, yet Fig. \ref{fig:3} clearly shows that the position of the CDW peak always coincides with the same fringe. 
This indicates that the CDW never completely vanishes and a residual characteristic of it survives the IR laser excitation even for timescales shorter than probed in our experiment. 

To interpret our findings it is instructive to consider the physical processes that occur in the nonequilibrium conditions shortly after the excitation. 
Therefore we have simulated the electron $T_e$ and lattice $T_l$ temperatures by solving (using the Euler method \cite{25}) the coupled differential equations within the two temperature model \cite{26}
\begin{equation}
\begin{split}
C_e(T_e)\frac{\mbox dT_e}{\mbox dt} &= -G (T_e-T_l) + S(t)\\
C_l(T_l)\frac{\mbox dT_l}{\mbox dt} &=  G (T_e-T_l),
\end{split}
\label{eq:1}
\end{equation}
where $G=4.6\cdot10^{11}$ W/(cm$^3$K) is the electron-phonon coupling constant \cite{27,28}, $C_e$=$\gamma T_e$ with $\gamma$=211 J/(m$^3$K) is the electron heat capacity \cite{28}, $C_l$ is the lattice heat capacity, which was calculated within the Debye approximation \cite{29}, $t$ denotes time, and $S(t)$ is the IR laser excitation, which was simulated as a Gaussian with 60 fs FWHM with a height adjusted to yield a final lattice temperature of 300 K. 
The sample is optically thin and homogeneously heated, thus the heat transport during the calculated time period can be neglected \cite{27}. 
Our calculations [Eq. \eqref{eq:1}] show the electron gas in Cr thin film is heated to about 2500 K (see inset in Fig. \ref{fig:4}(a)), well above the N\'{e}el temperature and the Fermi surface nesting condition is most likely vaporized shortly after the IR pulse excitation. 
Due to the electron-phonon coupling the electron gas cools down within 1.5 ps while heating the lattice to 300 K. 

%%%%%%%%%%%%%%%%%%%%%%%%%%%%%%%%%%%%%%%%%%%%%%%%%%%%%%%%%%%%%%%%%%%%%%%%%%%%%%%%%%%%%%%%%%
%%%%%%%%%%%%%%%%%%%%%%%%%%%%%%%%%%%%%%%%%%%%%%%%%%%%%%%%%%%%%%%%%%%%%%%%%%%%%%%%%%%%%%%%%%

Assuming the Fermi surface nesting is the origin for the CDW, intuitively one might expect that the CDW melts and the ions equilibrate to a displacement-free lattice. 
Yet our data shows that the CDW is never completely disrupted. 
We attribute the relatively slow dynamics of the periodic lattice distortion or strain wave to be the origin of the persistence of the CDW. 
The time where we expect the Fermi surface nesting condition to be absent (1.5 ps, see Fig. \ref{fig:4}(a)) is comparable to the reciprocal frequency $1/\nu$=1 ps for a longitudinal acoustic phonon with a wavelength corresponding to the CDW period \cite{30}. 
Therefore, the residual of this acoustic phonon may act as a seed for condensation of the charge ordered state. 
This observation is quite remarkable as here the emergence of the ordered electronic state appears to be driven by the structural lattice distortion, whereas typically the electronic states are the driving force \cite{4}.

In summary, we studied condensation dynamics of the structural order parameter in chromium. 
Using ultra-short laser pulses we excited the electronic subsystem and heated the sample close to the N\'{e}el temperature. 
Although we expect the electronic charge ordering to vanish shortly after excitation, we observe a strong CDW satellite peak persisting at the smallest measured time delay. 
The wavevector of the CDW reflection at elevated temperatures further indicates that the CDW never completely vanished, even for times shorter than the x-ray pulse duration. 
We find that after 100 ps the height of the satellite peak behaves in accordance with the equilibrium model. 
The persistence of the CDW in chromium is attributed to the long-lived lattice displacement associated with the CDW. 
The weakly first-order phase transition character at the N\'{e}el temperature and the pinning of the CDW at the film boundaries could also play an important role and further studies are desired to reveal dynamical interplay between electronic and ionic degrees of freedom, particularly at the ps time scales. 
Finally, we anticipate that the long-lived periodic lattice distortion and presumably lattice-induced ultrafast condensation of the electronic charge ordering can stimulate new concepts for ultra fast switching devices.

\acknowledgments
We acknowledge help of J. Wingert, J. Stanley, and T. Saerbeck in measuring the expansion coefficient of Cr. 
This work at UCSD was supported by U.S. Department of Energy, Office of Science, Office of Basic Energy Sciences, under Contract DE-SC0001805. 
This research used resources of the Advanced Photon Source, a U.S. Department of Energy (DOE) Office of Science User Facility operated for the DOE Office of Science by Argonne National Laboratory under Contract No. DE-AC02-06CH11357.

\bibliography{References.bib}

%%%%%%%%%%%%%%%%%%%%%%%%%%%%%%%%%%%%%%%%%%%%%%%%%%%%%%%%%%%%%%%%%%%%%%%%%%%%%%%%%%%%%%%%%%
%%%%%%%%%%%%%%%%%%%%%%%%%%%%%%%%%%%%%%%%%%%%%%%%%%%%%%%%%%%%%%%%%%%%%%%%%%%%%%%%%%%%%%%%%%
%appendix
%%%%%%%%%%%%%%%%%%%%%%%%%%%%%%%%%%%%%%%%%%%%%%%%%%%%%%%%%%%%%%%%%%%%%%%%%%%%%%%%%%%%%%%%%%
%%%%%%%%%%%%%%%%%%%%%%%%%%%%%%%%%%%%%%%%%%%%%%%%%%%%%%%%%%%%%%%%%%%%%%%%%%%%%%%%%%%%%%%%%%
\appendix
\renewcommand{\theequation}{A\arabic{equation}} 
\renewcommand{\thefigure}{A\arabic{figure}}
\setcounter{figure}{0}
\section{Interference between charge density wave satellites and the Bragg peak}
\label{A}
Here we aim to calculate the scattered intensity of a thin crystalline sample which contains a charge density wave (CDW). 
We in particular concentrate on the elastic component of the CDW, the periodic lattice displacement or strain wave (SW).
The CDW and SW have the same periodicity and are shifted by $\pi/2$ with respect to each other \cite{4}.
We restrict ourself to a thin homogeneous sample and consider $\mathbf r = (0,0,r)$ and $\mathbf q=(0,0,q)$, which is satisfied in our experiment.
The scattered amplitude  in the kinematic approximation can be written \cite{Warren1969}
\begin{equation}
F(q) = F_u(q) \sum_{n=0}^{N-1}e^{iqr_n},
\label{eq:A1}
\end{equation}
where $F_u(q)$ is the structure factor of the unit cell, that includes atomic scattering, and the summation is performed over the crystal layers.
In the presence of a SW the positions of the unit cells can be written as \cite{4}
\begin{equation}
r_n = r_n^0 +A_\textrm{SW}\cos(2Qr_n^0-\phi_0),
\label{eq:A2}
\end{equation}
where $r_n^0=n\cdot a$ are the positions without SW, $a$ is the lattice constant, $A_\textrm{SW}$ is the amplitude, $Q$ is the wave vector of the strain modulation, and $\phi_0$ defines the offset of the wave.
For $\phi_0=\pi/2$ we obtain a sine modulation with a node at the origin.
Substituting eq. \eqref{eq:A2} in \eqref{eq:A1}, using the well known expression \cite{Gradshteyn2000}
\begin{equation*}
e^{iz\cos(\phi)}=\sum_{k=-\infty}^{\infty}i^kJ_k(z)e^{ik\phi},
\end{equation*}
with $J_j$ being the Bessel function of $j$th kind,
and considering only terms of order $qA_\textrm{SW}$ yields ($qA_\textrm{SW}$ is typically on the order of $10^{-2}$ see Ref. \cite{4})
\begin{equation}
\begin{split}
F(q) &= F_u(q) \sum_{n=0}^{N-1}\left(e^{iqr_n^0}+\phantom{\frac12}\right.\\
&\left.+\frac{iqA_\textrm{SW}}{2}\left[e^{i\phi_0}e^{i(q-2Q)r_n^0}+e^{-i\phi_0}e^{i(q+2Q)r_n^0}\right]\right)
\end{split}
\label{eq:A3}
\end{equation}
All terms in eq. \eqref{eq:A3} can be readily evaluated \cite{Warren1969} and using
%\begin{equation*}
%\begin{split}
%F(q) =&  F_u(q)\cdot \left[f(q) \phantom{\frac12}\right.\\
%&\left.+ \frac{iqA_\textrm{SW}][2]e^{i\phi_0}}{2}f(q-2Q) + \frac{iqA_\textrm{SW}e^{-i\phi_0}}{2}f(q+2Q)\right],
%\end{split}
%\label{eq:A4}
%\end{equation*} 
$$f(q) = e^{i(N-1)qa/2}\frac{\sin(Nqa/2)}{\sin(qa/2)}.$$
the measured intensity can be written as
\begin{equation}
\begin{split}
I(q)	=&|F_u(q)|^2\left[|f(q)|^2\right.\\
&\left.-qA_\textrm{SW}\sin(\alpha)|f(q)f(q+2Q)|)\right.\\
&\left.+qA_\textrm{SW}\sin(\alpha)|f(q)f(q-2Q)|)\right.\\
&\left.+\frac{(qA_\textrm{SW})^2}4|f(q+2Q)|^2 + \frac{(qA_\textrm{SW})^2}4|f(q-2Q)|^2\right],
\end{split}
\label{eq:I}
\end{equation}
where $\alpha = Qa[N-1]-\phi_0$ and interference terms between the satellites $f(q-2Q)f(q+2Q)$ have been neglected, since $f(q)$ decays rapidly from its central peak.
% \begin{equation}
% \begin{split}
% I(q)	=&\left\vert F(q)\right\vert^2\\
% =&F_u(q)\left[|f(q)|^2-qA_\textrm{SW}\textrm{Re}(f(q)f^*(q-Q)) \right.\\
% &\left.+ qA_\textrm{SW}\textrm{Re}[f(q)f(q+Q)]\right].
% \end{split}
% \label{eq.I}
% \end{equation}
Bragg peaks are located at $q=G_{00l}=2\pi l/a$ with integer $l$ and the CDW peaks are located around the Bragg peaks $q = G_{00l}\pm Q$.

If the number of layers is large, then $f(q)$ can be well approximated by a Dirac $\delta$ function. 
In this case the interference terms can be neglected and the periodic charge ordering results in enhanced intensity at the satellite positions\cite{4}.
If the interference terms are present the intensity at the satellite positions can be increased or decreased dependent on the parameters.

In our measurement the number of layers is only 89 and the full expression \eqref{eq:I} has to be taken into account.
The fit using equation \eqref{eq:I} to the experimental data is shown in Figure 2 (a) of the main text.
The theoretical expression is increased at $q=G_{002}-2Q$ and decreased at $q=G_{002}+2Q$ .
For the obtained parameters we also find that the satellite intensity scales linear in $A_\textrm{SW}$ ($A_\textrm{SW}^2$ terms are about one order of magnitude smaller).

\begin{figure}[t]
\centering
\includegraphics[width=0.5\columnwidth]{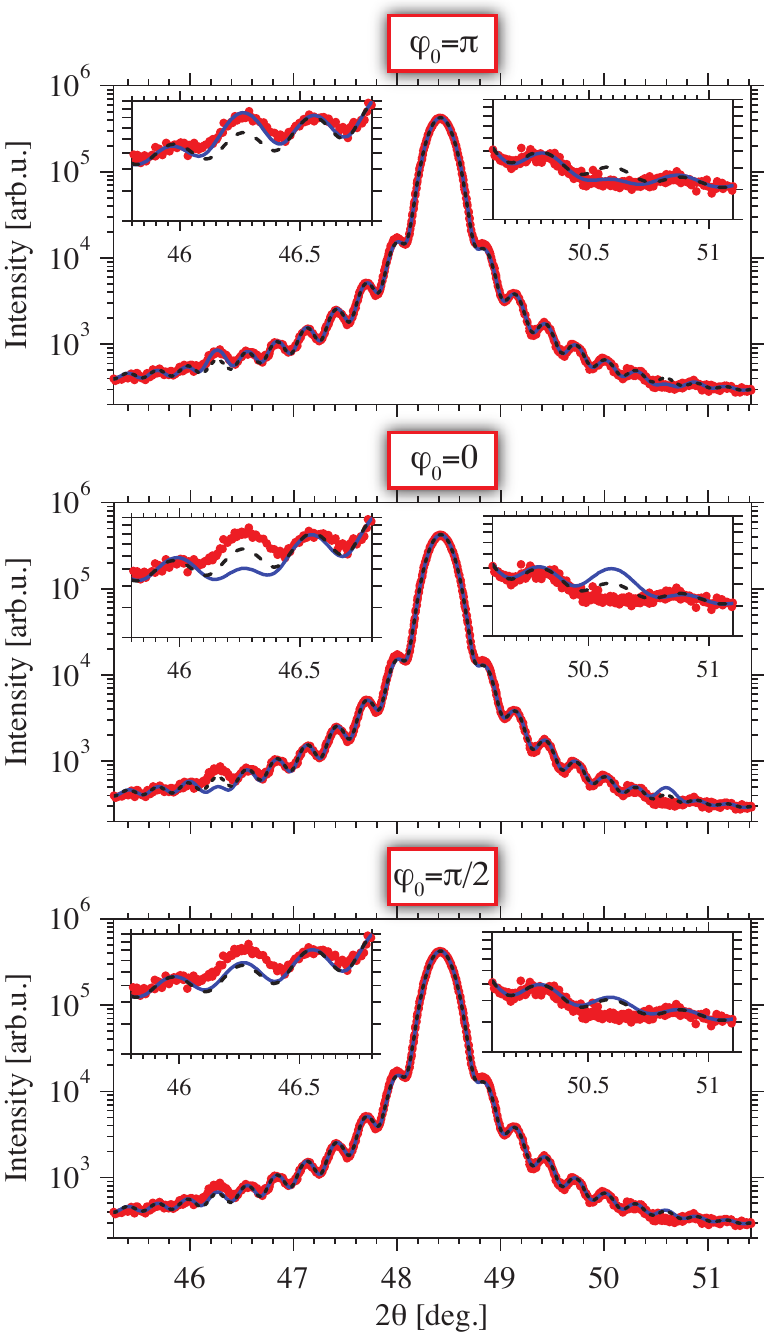}
\caption{X-ray diffraction scan of the Cr film cooled to 20 K (red points). A fit using equation \eqref{eq:I} assuming a displacement free lattice $A_\textrm{SW}=0$ (black dashed line) and a SW $A_\textrm{SW}>$ (blue solid line) with a varying phase of the SW with respect to the substrate-Cr interface. The amplitude and period of the SW are fixed. The data is consistent with $\phi_0=\pi$, which is described in the main text (see Figure 2a of the main text).}
\label{fig:cossin}
\end{figure}
\section{Fitting of the diffraction data}
\label{B}
We have modeled the data by fitting eq. \eqref{eq:I} to it.
To determine the film properties expected without the presence of a SW, two regions (3 fringes wide) centered at both SW satellites were excluded from the initial fit and eq. \eqref{eq:I} with $A_\textrm{SW}=0$ was used. 
Equation \eqref{eq:I} was convolved with a effective instrument function, accounting for mosaicity and x-ray bandwidth.
Further the fit was multiplied by an exponential function, taking into account the Debye-Waller factor, and was multiplied by a surface roughness term of the form $\exp(-\sigma^h [q-q_0]^h)$, where $q_0$ is the position of the Bragg peak. Interestingly, $h=1.2$ gave the best fit. 
A linear background term was subtracted. 
A thickness of 25.6$\pm$0.3 nm was obtained in this fit.
The surface roughness is below 1 nm, which manifests itself in well resolved fringes out to the maximum measured q-range corresponding to about 2.5 nm.

To determine the SW amplitude the data in the regions previously excluded (3 fringes around each SW satellite peak) was fit  by equation \eqref{eq:I}, now with $A_\textrm{SW}>0$.
Initially we fit the wave vector $Q$ of the SW and confirmed the SW has an integer number of half periods within the film. 
Then the wave vector was fixed and only the amplitude $A_\textrm{SW}$ and phase $\phi_0$ of the SW were fit here, while keeping all other operations the same as in the previous fitw. 
During this procedure a SW amplitude, $A_\textrm{SW}=0.5\pm0.2$\% of the lattice parameter $a=0.29~nm$, and phase $\phi_0=\pi\pm1$ were found.  
The goodness of fit was characterized by $R=(\sum_i[I^\textrm{th}_i-I^\textrm{exp}_i]^2)/\sum_i[I^\textrm{exp}_i]^2$ and the confidence intervals ware determined from the values for which $R$ was twice as large as the minimum value, while all other fit parameters were fixed.

Figure \ref{fig:cossin} illustrates the interference effect and the sensitivity to the phase of the SW with respect to the film boundaries. 
Three fits with the same amplitude and period of the SW, but with different phases are shown: $\phi_0=\pi$, $\phi_0=\pi/2$, and $\phi_0=0$. 
Clearly, the phase of the SW has a large  impact on the diffraction signal. 
For instance for $\phi_0=0$ and $\phi_0=\pi/2$ , which correspond to positive displacement and a node at the substrate-Cr interface, respectively, the simulated diffraction signal is significantly different from the measured diffraction signal. 
For $\phi_0=0$ the enhancement of the intensity occurs for $q$-values higher than the scattering vector of the Bragg peak, contrary to the observations during our experiment. 
For $\phi_0=\pi/2$ both peaks are symmetric, which also fails to reproduce the measured data.
The simulation with $\phi=\pi$ shown in the main text agrees well with the measured data. 

\section{Scanning electron microscopy image of the sample}
\label{C}
\begin{figure}[h]
\centering
\includegraphics[width=0.4\columnwidth]{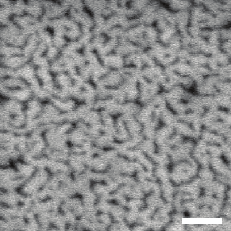}
\caption{Scanning electron microscopy (SEM) image of the sample. The scale bar shows 200 nm.}
\label{fig:SEM}
\end{figure}
A scanning electron microscopy image of the surface of the thin film is presented in Figure \ref{fig:SEM}. 
The electron beam energy was 6 keV. 
\section{Theoretical fitting of the film temperature as a function of the time delay}
\label{D}
\begin{figure}[t]
\centering
\includegraphics[width=0.5\columnwidth]{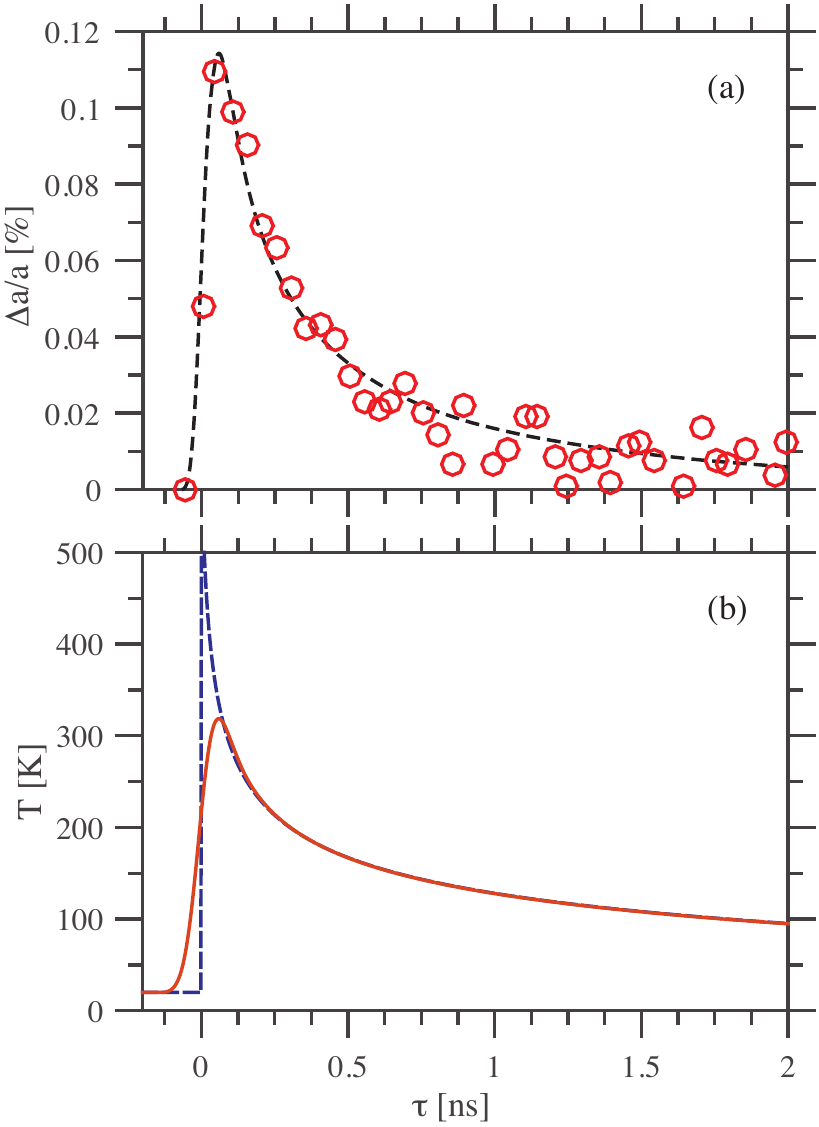}
\caption{(a) The measured lattice expansion (red circles) and the simulation using equation \eqref{eq:Kapitza} (black dashed line). 
(b) The calculated temperature of the film before (red solid line) and after convolution with a Gaussian function 100 ps FWHM (blue dashed line), representing smearing by the x-ray pulse. }
\label{fig:S3}
\end{figure}

We have used the following equation to model the film temperature decrease \cite{Stoner1993}
\begin{equation}
d\cdot C(T_f)\frac{\mbox d T_f}{\mbox dt} = - \sigma_K\cdot (T_f-T_s),
\label{eq:Kapitza}
\end{equation}
where $T_f$  and $T_s$ are the temperatures of the film and the substrate, $\sigma_K$ is the Kapitza interface conductance constant, $C(T_f)$ is the specific heat and $d$ is the thickness of the film.
Due to a high thermal conductance of MgO we considered $T_s=20$ K as a constant.
The Kapitza constant was modeled cubic following Ref. \cite{Young1989} and the heat capacity of the film was modeled within the Debye approximation, as in the main text (the contribution of the electronic system is neglected here).
A Kapitza constant of $\sigma_K=\alpha(T)^3$ with $\alpha=10.5\pm2$ W/(m$^2$K$^4$) and a maximum film temperature of $T_f^\textrm{max}=580\pm200$ K were determined by fitting the temperature curves obtained from \eqref{eq:Kapitza} to the data presented in Figure 4(a) of the main text. 
The temperature of the film was smoothen by a resolution function with a width of 100 ps, being FWHM of the x-ray pulses. 
Figure \ref{fig:S3} (a) shows the relative lattice expansion for all measured values up to 2 ns. 
Figure \ref{fig:S3} (b) presents the calculated temperature with and without the convolution with the Gaussian x-ray pulse shape. 

\end{document}